\documentclass[article,twocolumn]{aastex631}
\def\beq{\begin{equation}}
\def\eeq{\end{equation}}
\def\I{{\sc i}}

\usepackage{natbib}

\begin{document}


\title{A revised comparison of distant and nearby solar twins}


\author{Charles R. Cowley}
\affiliation{Department of Astronomy, University of Michigan, 
1085 S. University, Ann Arbor, MI 481090-1107\\
orcid:0000-0001-9837-3662}                            

\author{Robert E. Stencel}
\affiliation{Chamberlin Observatory, University of Denver, 2930 E Warren Ave., 
     Denver, CO 80210, USA, \\
     orcid:0000-0001-8217-9435\\
     e-mail: robert.stencel@du.edu}

\begin{abstract} Properties of solar twins \citep{leh23} at kiloparsec distances from the 
local standard of rest (LSR) are compared to solar twins within 100 pc of the Sun. 
They have velocity distributions closely similar to those of the nearby twins, in addition to closely matching $T_{\rm eff}$, $\log{(g)}$ and $[Fe/H]$.  The new twins are at slightly higher galactic latitudes, and are somewhat closer to the Galactic center.  Additionally, they may be significantly older than nearby solar twins.
\end{abstract}

\large

\section{Distant solar twins}
Solar twins have been discussed extensively in the literature 
\citep[see][henceforth, BD18]{bd18}.
The BD18 stars are all within 100 pc of the Sun.  In this note we discuss 
potential distant twins or analogues
as investigated by \citet{leh22,leh23}  
and \citet{liu22}. 
The authors designate their work 
SDST (Survey for Distant Solar Twins). 
Twins and analogues are defined in their work by
limits on effective temperature, surface gravity, and [Fe/H],
\footnote{We use the notation
$[Y/X]$ = $\log(Y/X)_{\rm star}-\log(Y/X)_{\rm Sun}$} 
with kiloparsec distances--far
greater than for twins previously investigated.

A goal of SDST was to 
provide observations that might detect variations in the fine structure constant over kiloparsec
distances. \citet{mur22} give an introduction to the problem of fine structure variation 
and the proposed investigation method using sunlike spectra.

SDST explored   
one sharply limited direction of the Galaxy, an area of only 3.02 square degrees, in
the direction of Galactic longitude $330.0$ deg, latitude 16.55 deg.
The survey yielded 877 candidate stars
with Gaia DR3 distances $483 \le d \le 5425$ (mean=1933) pc.   
Twenty of the 877 stars were designated as twins
and 299 as solar analogues.  

\section{Sightline}
Figure 1 is a plot of $Z$-values of the 877 SDST stars
vs. $\rho$ = $\sqrt(X^2 + Y^2)$, where the origin of the coordinates is located at the mean sun position.  The line of sight is out of the plane, the most distant of the 877
stars is Z = 1495 pc.  The $\rho$ values of the 20 solar twins are indicated by the 
 blue arrows.  This sightline encroaches on the so-called thick disk.  The idea of a thick galactic disk was proposed by \citet{ab03} -- see also \citet{pra23} -- to describe `accretion debris' from galactic mergers.   A scale height of 1.5 kpc would be seen at a distance of 4.5 kpc, for objects observed at galactic latitude 16 degrees.  The Lehmann sample includes a few objects between 4.5 and 5.5 kpc distance and thus, could include thick disk components with potentially altered composition. 
 
\begin{figure}
 \includegraphics[width=\columnwidth]{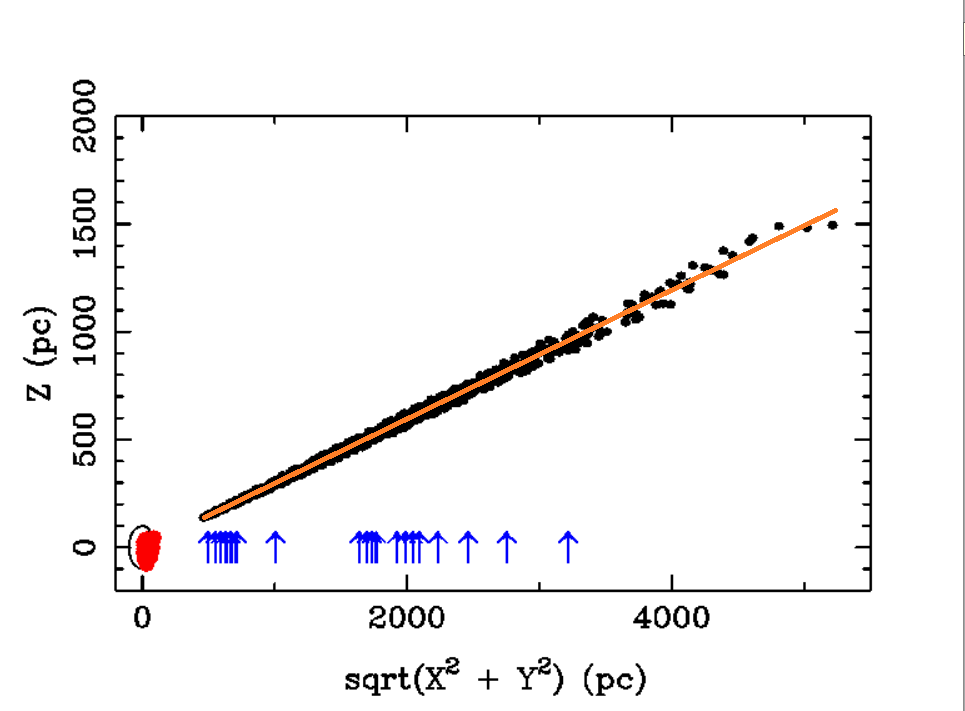}
 \caption{ Z-values vs. $ \sqrt(X^2 + Y^2)$ for 877 SDST stars (black filled circles).  
 The red clump in the lower left corner are the 79 BD18 stars, mostly with velocities less than 63 km/sec.   The orange line through the
 SDST points shows their allignment which departs very slightly for the
 most distant stars. See text for further discussion. }
\label{fig:Fig1}  
\end{figure}

The vertical scatter in Figure 1 is partially due to small differences in the line
of sight to the SDST stars, and partially due to errors in the 
Gaia parallaxes ($\varpi$ and $\Delta\varpi$).  It is
beyond the scope of this paper to explore the latter in detail \citep[see][]{lur18}.  A
naive interpretation of the distance errors, however, between 
1/($\varpi$ + $\Delta\varpi$) and
1/($\varpi$ - $\Delta\varpi$) would lead to far greater scatter than is seen in 
Figure 1.

 \section{Kinematics: velocities}

We have computed the velocities of the 20 twins and all 877 of the SDST stars, 
and compared them 
with the kinematical properties of sets of solar twins discussed by \citet[henceforth, CY22]{CY22} 
notably, 79 stars with precision differential abundances from BD18,  
which we consider definitive twins for the local solar neighborhood.
Because the SDST twins are in a different region of the Galaxy, and
at significantly greater distances than those of the
BD18 sample, we focus on the velocities relative to the LSR;
the distribution of stars in physical space is distinct by definition
(see Figure 2).

\begin{figure}
 \includegraphics[width=\columnwidth]{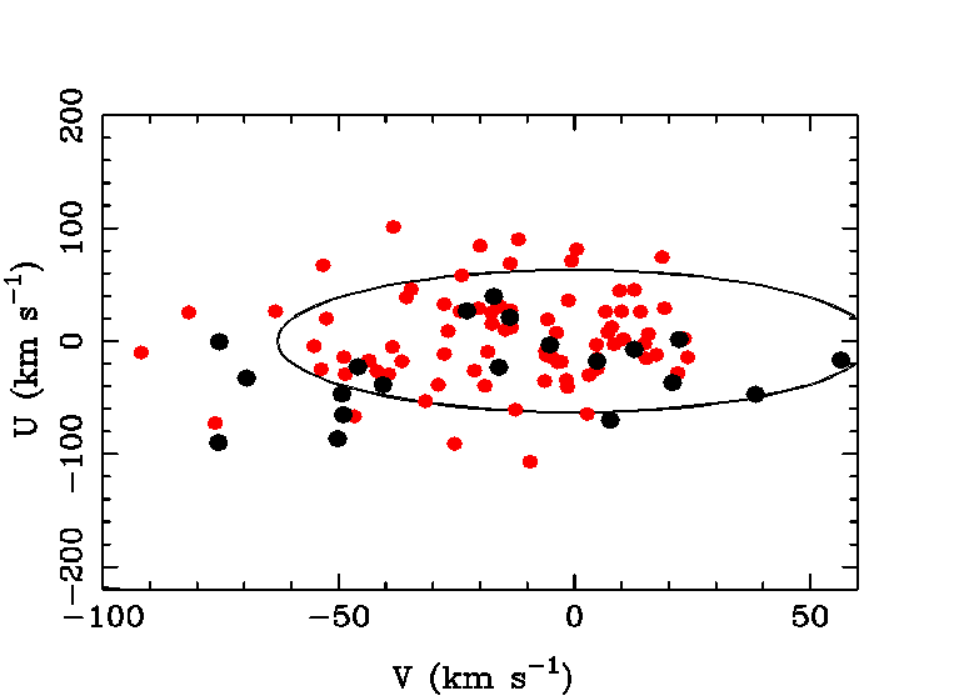}
 \caption{Velocities with respect to the LSR of solar twins, near 
 (BD18, red) and distant
 (SDST, black) from the solar neighborhood, showing little difference in the distribution
 of the two sets of stars.  
 The circle has a radius of 63 km sec$^{-1}$.  It
 is distorted because of the different $U$ and $V$ scales.
\label{fig:Fig2}}
\end{figure}
We also consider the velocities in the Z-direction with the help of Figure 3.
\begin{figure}
 \includegraphics[width=\columnwidth]{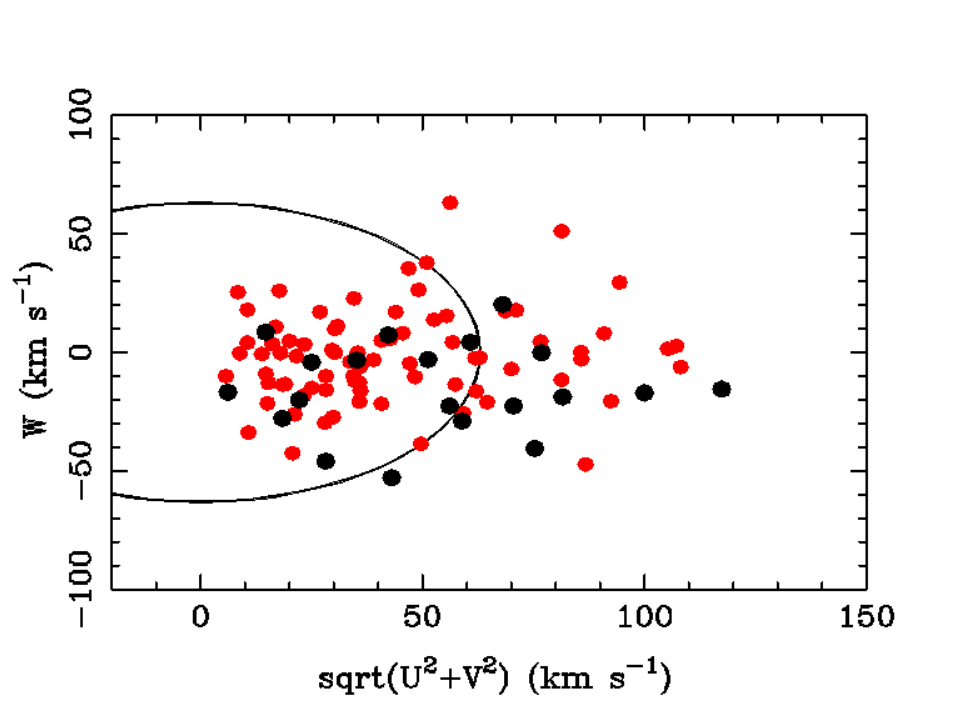}
 \caption{Velocities with respect to the LSR in the Z and XY directions.  The
 similarity of the BD18 (red) and SDST (black) distributions is clear.
}
\label{fig:Fig3}
\end{figure}

Finally, it is of interest to examine the distributions of velocity in the 
phase space of kinetic energy (per unit mass) vs. $L_Z$, the Z-component of the angular
momentum (per unit mass).  The solid line is part of a parabola that describes the locus of
circular orbits near the Sun.  Both the BD18 and SDST points adhere closely
to that line. 

\begin{figure}
 \includegraphics[width=\columnwidth]{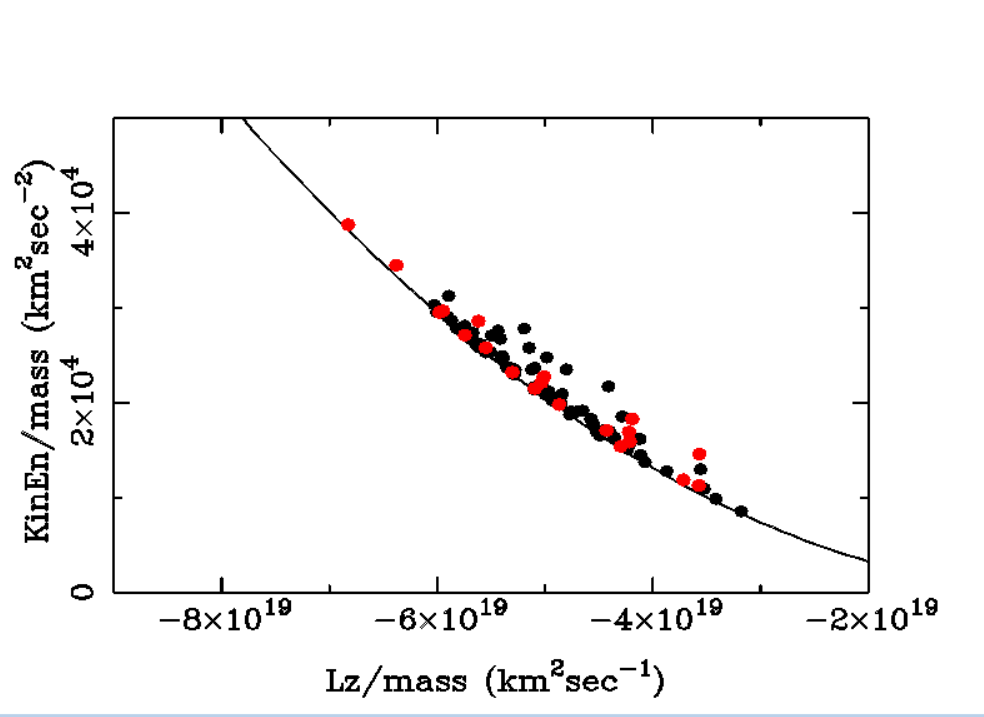}
 \caption{BD18 and SDST velocities in phase space.
}
\label{fig:Fig4}
\end{figure}

We conclude that the velocity kinematics of the 20 SDST distant twins are
closely similar to BD18 objects. 

\section{ages}
Both the BD18 and SDST ages were computed using the q2 algorithm of 
\citet{ram14}.  The mean age differences are
of interest.  For 12 twins from the BD18 stars, CY22 found a mean age of
4.86 Gyr (see their Table 2).  In CY22, four other sources of 12 twins were discussed, with
mean ages computed for this paper.  These ages are:
4.01 \citep{dm19}, 4.42 \citep{liu22}
4.78 \citep{brew16}, and 3.96 Gyr \citep{niss15}.
The average of these five means is 4.41 Gyr.  We take 1.5 Gyr as a 
working, conservative uncertainty of an 
individual twin's age--roughly a standard deviation (see \citet{dm19} and CY22).  
Dividing these N - 1 values by $\sqrt(11) = 3.32$,
these give an uncertainty of 0.45 Gyr for the mean. Altogether, we may say that solar twins in
the solar neighborhood have a mean age of $4.43 \pm 0.45$ Gyr.
 
The mean distance of the 20 distant twins from the Galactic center is $6.72 \pm 0.65$ kpc.  They 
have a mean age of 8.35 Gyr and if we again use 1.5 Gyr as a typical
uncertainty, the uncertainty of this mean is $1.5 \sqrt(19)=0.34$ Gyr. So the distant twins
have a mean age of $8.35 \pm 0.34$ Gyr.  The mean age of the near twins plus 2 sigma is 5.25 Gyr
while the mean age of the distant twins minus 2 sigma is 7.69, leaving a substantial gap.

Judging by conservative estimates for the random
errors, the distant twins are significantly ($> 2\sigma$) older than the nearby ones.
However, the typically larger {\em systematic} errors could change that conclusion.  
  
\section{Summary}
The distant solar twins of the SDST studies have very similar velocity distributions to
those of the classic BD18 stars in the solar neighborhood, supporting their validity 
as solar twins, and supporting their use in the investigation of possible variations
in the fine structure constant.  An age difference in the near and distant twins is of interest
because of a possible connection with galactic chemical evolution, and/or violation of the Vogt-Russell theorem \citep{rus31}.


\begin{thebibliography}

\bibitem[Abadi et al.(2003)]{ab03} Abadi, M., Navarro, J., Steinmetz, M., Eke, V. 2003 ApJ 591, 499. doi:10.1086/375512 

\bibitem[Bedell, et al.(2018)]{bd18} Bedell M., Bean J.~L., Mel{\'e}ndez J.,  
et al., 2018 ApJ 865, 68. doi:10.3847/1538-4357/aad908(BD18) 

\bibitem[Brewer et al.(2016)]{brew16} Brewer, J.~M., Fischer, D.~A., Valenti, J.~A., 
et al.\ 2016 \apjs 225 32. doi:10.3847/0067-0049/225/2/32 (BR)


\bibitem[Cowley \& Y{\"u}ce(2022)]{CY22} Cowley C.~R., Y{\"u}ce K., 
2022 MNRAS 512, 3684. doi:10.1093/mnras/stac637 (CY22)


\bibitem[Delgado Mena et al.(2019)]{dm19} Delgado Mena E., Moya A., 
Adibekyan V.,  et al., 2019, A\&A, 624, A78. doi:10.1051/0004-6361/201834783

\bibitem[Lehmann et al.(2022)]{leh22} Lehmann, C., Murphy, M.~T., Liu, F., et al.\ 
2022 \mnras  512, 11. doi:10.1093/mnras/stac421 (SDST I)
   
\bibitem[Lehmann et al.(2023)]{leh23} Lehmann, C., Murphy, M.~T., 
Liu, F., et al.\ 2023 \mnras  521, 148. doi:10.1093/mnras/stad381 (SDST III)

\bibitem[Liu et al.(2022)]{liu22} Liu, F., Murphy, M.~T., Lehmann, C., et al.\ 
2022 \mnras 517, 5569. doi:10.1093/mnras/stac3033

\bibitem[Luri et al.(2018)]{lur18} Luri, X., Brown, A.~G.~A., 
Sarro, L.~M., et al.\ 2018 \aap 616, A9. doi:10.1051/0004-6361/201832964

\bibitem[Murphy et al.(2022)]{mur22} Murphy, M.~T., Berke, D.~A., 
Liu, F., et al.\ 2022 Science  378, 634. doi:10.1126/science.abi9232   

\bibitem[Nissen(2015)]{niss15} Nissen, P.~E.\ 2015 \aap 579, A52. doi:10.1051/0004-
6361/201526269

\bibitem[Prantzos et al. (2023)]{pra23} Prantzos, N., Abia, C., Chen, T. et al. \ 2023 
https://arxiv.org/pdf/2305.13431.pdf 

\bibitem[Ram{\'\i}rez et al.(2014)]{ram14} Ram{\'\i}rez, I., 
Mel{\'e}ndez, J., Bean, J., et al.\ 2014 \aap 572, A48. doi:10.1051/0004-6361/201424244

\bibitem[Russell (1931)]{rus31} Russell, H.N. \ 1931 MNRAS 91, 951, doi.org/10.1093/mnras/91.9.951. 

\bibitem[Spina et al.(2018)]{spin18} Spina L., Mel{\'e}ndez J., Karakas A.~I.,  
et al., 2018 MNRAS 474, 
2580. doi:10.1093/mnras/stx2938

\end{thebibliography}
\end{document}